\documentclass[letterpaper]{article} 
\usepackage{aaai2026}  
\usepackage{times}  
\usepackage{helvet}  
\usepackage{courier}  
\usepackage[hyphens]{url}  
\usepackage{graphicx} 
\usepackage{natbib}  
\usepackage{caption} 
\usepackage{subfig} 
\usepackage{xcolor}
\usepackage{fontawesome} 
\usepackage{pifont}
\usepackage{natbib}
\usepackage{amsmath}
\usepackage{amssymb}
\usepackage{booktabs}
\usepackage{algorithm}
\usepackage{algorithmic}

\usepackage{newfloat}
\usepackage{listings}
\DeclareCaptionStyle{ruled}{labelfont=normalfont,labelsep=colon,strut=off} 
\floatstyle{ruled}
\newfloat{listing}{tb}{lst}{}
\floatname{listing}{Listing}
\title{Detect All-Type Deepfake Audio: Wavelet Prompt Tuning for Enhanced Auditory Perception}
\author{
	Yuankun Xie\textsuperscript{\rm 1},
	Ruibo Fu\textsuperscript{\rm 2}$^{*}$,
	Xiaopeng Wang\textsuperscript{\rm 3},
	Zhiyong Wang\textsuperscript{\rm 3},
	Songjun Cao\textsuperscript{\rm 4},
	Long Ma\textsuperscript{\rm 4}, \\
	Haonan Cheng\textsuperscript{\rm 1},
	Long Ye\textsuperscript{\rm 1}\thanks{Corresponding author}
}

\affiliations{
	\textsuperscript{\rm 1}State Key Laboratory of Media Convergence and Communication, Communication University of China, Beijing, China\\
	\textsuperscript{\rm 2}Institute of Automation, Chinese Academy of Sciences, Beijing, China\\
	\textsuperscript{\rm 3}School of Artificial Intelligence, University of Chinese Academy of Sciences, Beijing, China\\
	\textsuperscript{\rm 4}YouTu Lab, Tencent, Beijing, China\\
	xieyuankun@cuc.edu.cn, ruibo.fu@nlpr.ia.ac.cn, yelong@cuc.edu.cn
}

\usepackage{bibentry}

\begin{document}
	
\maketitle

\begin{abstract}
	The rapid advancement of audio generation technologies has escalated the risks of malicious deepfake audio across speech, sound, singing voice, and music, threatening multimedia security and trust. While existing countermeasures (CMs) perform well in single-type audio deepfake detection (ADD), their performance declines in cross-type scenarios. This paper is dedicated to studying the all-type ADD task. We are the first to comprehensively establish an all-type ADD benchmark to evaluate current CMs, incorporating cross-type deepfake detection across speech, sound, singing voice, and music. Then, we introduce the prompt tuning self-supervised learning (PT-SSL) training paradigm, which optimizes SSL front-end by learning specialized prompt tokens for ADD, requiring 458× fewer trainable parameters than fine-tuning (FT). Considering the auditory perception of different audio types, we propose the wavelet prompt tuning (WPT)-SSL method to capture type-invariant auditory deepfake information from the frequency domain without requiring additional training parameters, thereby enhancing performance over FT in the all-type ADD task. To achieve an universally CM, we utilize all types of deepfake audio for co-training. Experimental results demonstrate that WPT-XLSR-AASIST achieved the best performance, with an average EER of 3.58\% across all evaluation sets. 
\end{abstract}

\section{Introduction}
With the development of audio language model (ALM) technology, it has become increasingly easy to synthesize any type of audio, including deepfake speech, sound, singing voice, and music. These deepfake audios pose a threat to society in various fields such as media, entertainment, cybersecurity, and political communication. 
Fortunately, research on audio deepfake detection (ADD) has been increasing annually. Among these, the earliest studies focused on deepfake speech detection. Researchers have developed a series of deepfake countermeasures (CMs) aimed at effectively detecting deepfake speech, based on the ASVspoof challenges \cite{todisco19_interspeech, liu2023asvspoof, wang2024asvspoof5}. Currently, some ADD research has gone beyond speech, such as the detection of deepfake singing voices \cite{zhang2024svdd, xie2024fsd, zang2024ctrsvdd}, sounds \cite{xie2024fakesound, xie2025codecfake}, and music \cite{comanducci2024fakemusiccaps}.

\begin{figure}[!t]
	\centering
	\subfloat{\includegraphics[width=3.3in]{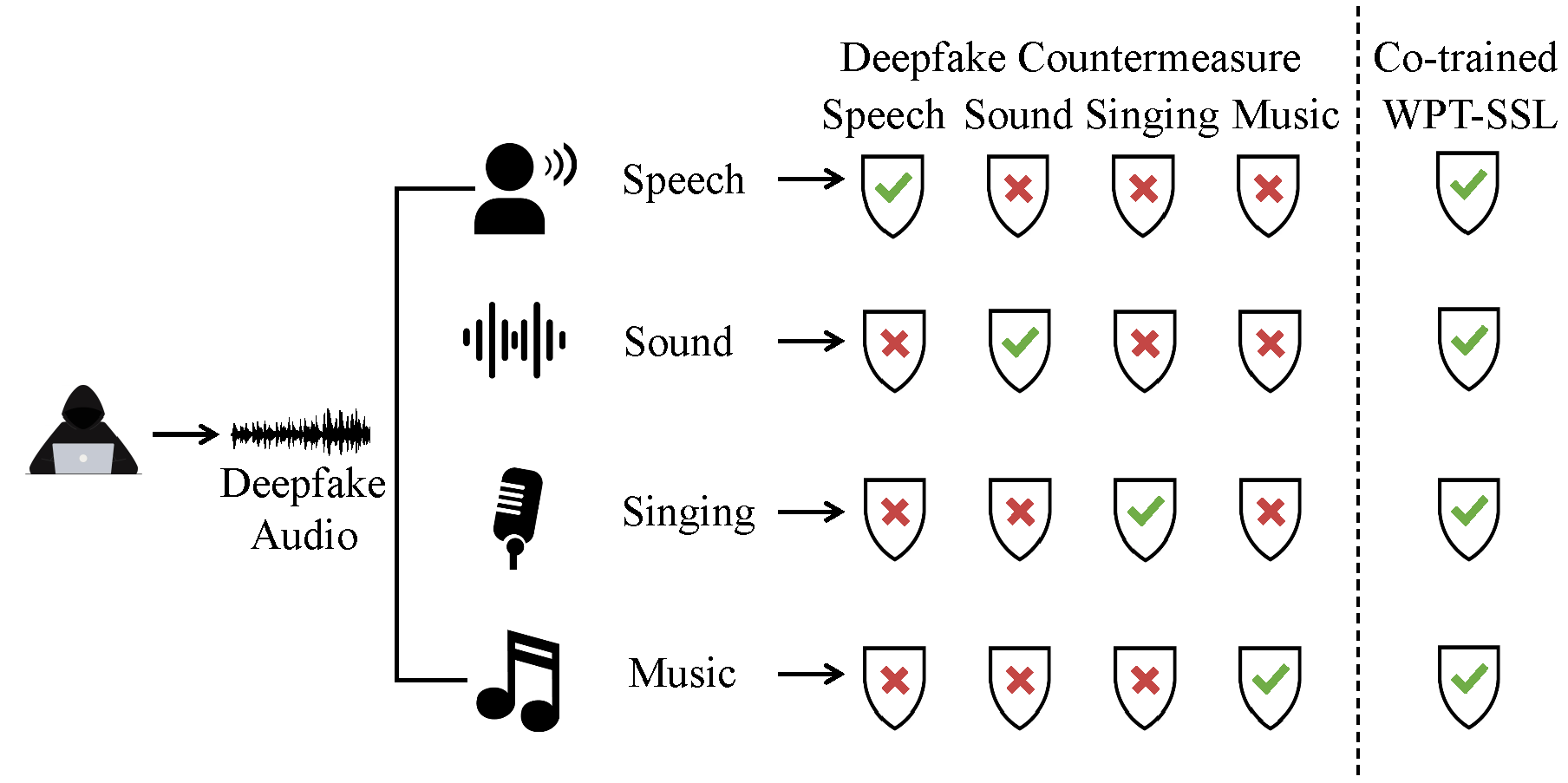}}
	\hfil
	\caption{The challenge for current single-type trained CMs toward cross-type ADD task, highlighting the effectiveness of our proposed WPT-SSL CM.}
	\label{fig:problem} 
\end{figure}

Although each type of deepfake audio has its corresponding countermeasure (CM), in real-world scenarios, the type of audio is often uncertain and may encompass one or more categories, such as speech, sound, singing voice, or music. This leads to the challenge that the CM trained on a single type being unable to generalize and detect all types of audio, as shown in Figure \ref{fig:problem}. Therefore, it is crucial to develop an advanced CM that can generalize and effectively detect all-type of deepfake audio.

For CMs, the most effective approach currently is to use pre-trained self-supervised learning (SSL) features along with a classification backbone. A representative CM in speech deepfake detection is XLSR-AASIST \cite{tak2022automatic}, which fine-tunes (FT) the wav2vec2-xls-r (XLSR) \cite{babu22_interspeech} model on speech deepfake detection dataset, achieving excellent intra-domain (ID) and out-of-domain (OOD) generalization performance. However, when dealing with the all-type ADD task, several challenges are encountered. Firstly, from the data perspective, it is uncertain whether a CM trained on single audio type can generalize to detect other types of deepfake audio. Although some studies have investigated cross-type detection for two types \cite{gohari2025audio, xie2025codecfake}, there has been no exploration of cross-type detection for all audio types. Secondly, there has been no investigation into whether a domain-invariant feature exists that can ensure the invariance of authenticity discrimination across different audio types. This requires a detailed investigation of various SSL features as well as handcrafted features. Lastly, concerning the algorithm, although fine-tuning can yield promising results, it is highly dependent on specific hyper-parameters and requires a significant amount of training parameters \cite{wang2025mixture}. 

To address the aforementioned challenges, in this paper, we aim to develop an all-type audio deepfake CM. We are the first to comprehensively establish an all-type ADD benchmark, which includes cross-type deepfake detection among speech, sound, singing voice, and music. For the feature of CMs, we investigate handcrafted features, raw waveforms, and various SSL-based features through both freezing and fine-tuning. For the back-end classifier, we use AASIST \cite{jung2022aasist}, the most popular model in the field of ADD, as the back-end, and combine it with SSL front-end to form SSL-AASIST.

To efficiently optimize SSL front-end, inspired by Visual Prompt Tuning (VPT) \cite{jia2022visual}, we proposed the Prompt Tuning (PT)-SSL training paradigm for ADD task. PT-SSL introduces learnable prompt tokens before the input of each transformer layer, while keeping the other parameters of the layers frozen, with the goal of learning specialized prompt tokens for the ADD task. Furthermore, considering the human perception of different audio types, the primary differences in perceiving audio types lie in their frequency domain distributions \cite{norman2015distinct, kell2018task, munkong2008auditory}. However, current SSL models like wav2vec2, which are primarily designed for speech recognition, focus on temporal and specific speech frequency information, lacking the ability to capture full-frequency information. To enhance frequency domain adaptability and enable SSL-based CM to quickly adapt to all types of deepfake audio, we propose wavelet prompt learning (WPT)-SSL method. WPT-SSL applies a discrete wavelet transform (DWT) to a portion of the prompt tokens, obtaining tokens for different frequency bands, thereby enhancing the full-frequency perception capability of SSL-based CM. Surprisingly, we discovered that WPT-SSL can learn a type-invariant deepfake detection prompt in a specific frequency band (HH) obtained through wavelet decomposition, thereby enabling all-type audio deepfake detection.

We summarize the contributions of this work as follow:
\begin{itemize}
	\item{We proposed all-type ADD task and established a comprehensive benchmark to measure the current CM's capability in detecting all-type deepfake audio.}
	\item{To efficiently train SSL front-end, we proposed the PT-SSL training paradigm, which significantly reduces the number of training parameters by only learning prompt tokens, achieving performance close to FT.}
	\item{Considering the human perception of different audio types, we proposed the WPT-SSL method, which can learn type-invariant frequency authenticity information. Without adding extra training parameters, WPT outperformed FT under all ADD test conditions.}
	\item{To achieve an universally CM, we utilize all types of deepfake audio for co-training. Experimental results demonstrate that WPT-SSL-AASIST achieved the best performance with an average EER of 3.58\%.}
\end{itemize}

\begin{figure*}[!t]
	\centering
	\subfloat{\includegraphics[width=5in]{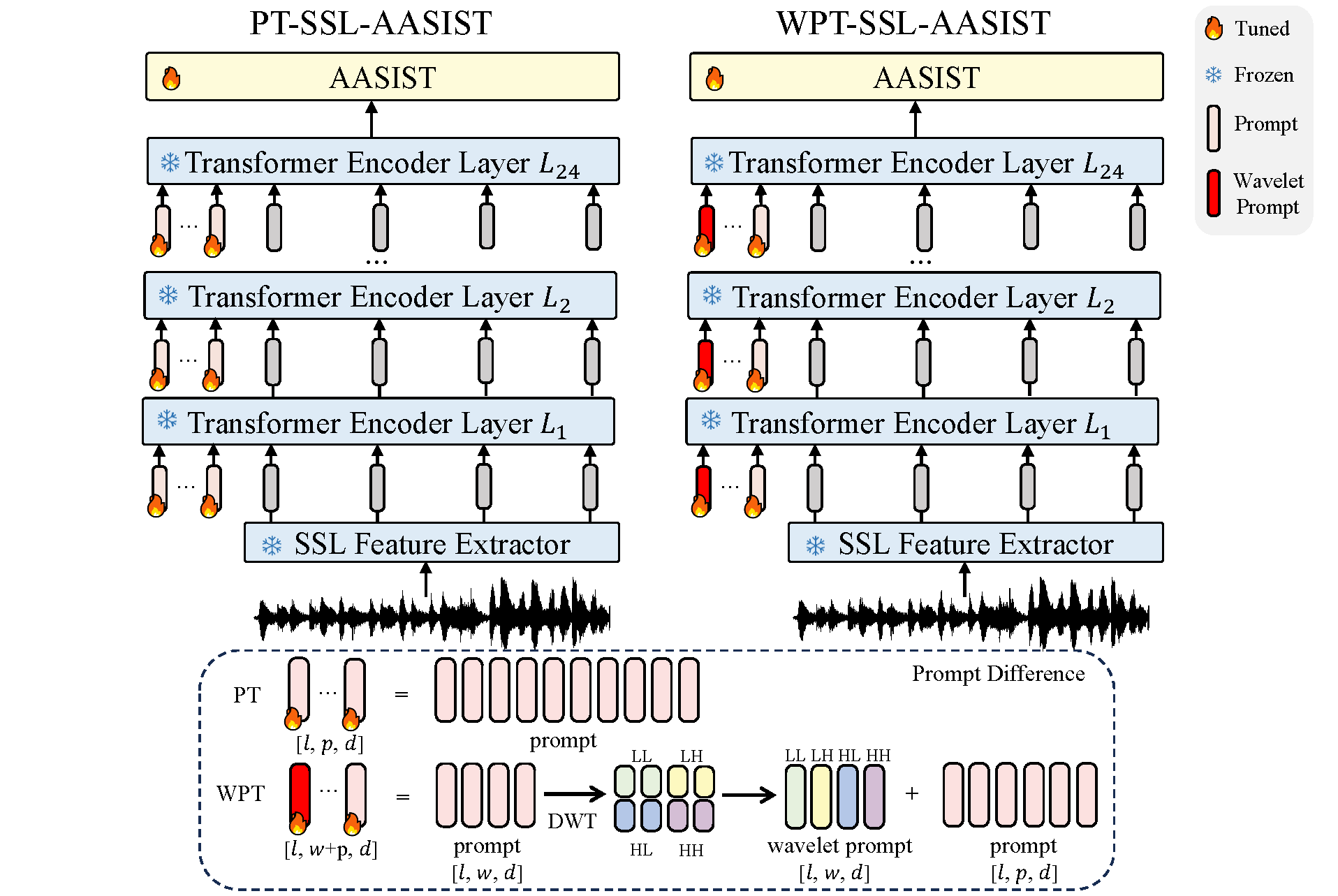}}
	\hfil
	\caption{Our proposed PT-SSL-AASIST (left) and WPT-SSL-AASIST (right). The differences between PT and WPT are illustrated below. WPT enhances the full-frequency perception of SSL-AASIST by applying DWT to part of the prompt tokens.}
	\label{fig:pipeline} 
\end{figure*}

\section{Prompt Tuning Countermeasure}
\label{sec:cm}
In this section, we introduce our proposed PT-SSL-AASIST and WPT-SSL-AASIST paradigm, which rapidly adapt SSL features to the ADD task by learning prompt tokens.
\subsection{PT-SSL-AASIST}
For an input audio $X$, we first pad or chop it to a fixed length $L$, obtaining the audio input $X \in \mathbb{R}^{L}$. Then, the audio input is first passed through the frozen SSL front-end feature extractor. For SSL implementations such as XLSR, this feature extractor comprises a 7-layer CNNs. Subsequently, we obtain the input to the first encoder layer of the transformer, $E_{0} \in \mathbb{R}^{t\times d} $, where $t$ represents the temporal length of the audio sequence and $d$ denotes the dimension of the transformer hidden states.
For the prompt token, we employ Xavier uniform initialization for all layers, resulting in $\mathbf{P}=\left\{\mathbf{P}_{k} \in \mathbb{R}^{p \times d} \mid k \in \mathbb{N}, 1 \leq k \leq l\right\}$, where $l$ represents the number of SSL layers, $p$ denotes the preset number of tokens for PT. Therefore, the input and output of the first layer of the Transformer are as follows:
\begin{equation}
	[Z_{1}, E_{1}] = \textcolor{blue}{L_{1}}([\textcolor{red}{P_{1}},E_{0}]),
\end{equation}
where $Z_{1} \in \mathbb{R}^{p\times d} $ is the variable generated by the first frozen transformer encoder at the prompt token position, which will be replaced by $P_{1} \in \mathbb{R}^{p\times d} $in the next computation. Thus, the PT calculation for other layers is as follows:
\begin{equation}
	[Z_{i}, E_{i}] = \textcolor{blue}{L_{i}}([\textcolor{red}{P_{i}},E_{i-1}]),     \quad \text{for } i=2,3,...,l.
\end{equation}
Taking the most commonly used SSL feature in the ADD domain, XLSR-300m, as an example, after the final 24 layers, we obtain a matrix output $I = [Z_{24}, E_{24}]$. $I$ will serve as the input to AASIST. The back-end AASIST classifier fully follows the structure of SSL-AASIST by Tak et al. \cite{tak2022automatic}, utilizing spectro-temporal graph attention to capture time-frequency features. The final output is a two-dimensional logits score, which is optimized through weighted cross-entropy (WCE) loss.

\subsection{WPT-SSL-AASIST}
In the PT-SSL-AASIST framework, the initial embedding $E_{0} \in \mathbb{R}^{t\times d} $, extracted by the SSL front-end, retains high temporal resolution from raw waveform inputs but lacks explicit frequency distribution and cross-type frequency attention ability. To achieve frequency-sensitive modeling for all-type ADD, we proposed WPT-SSL-AASIST, which introduces wavelet prompt tokens to enhance the frequency perception capability of SSL.
The difference between WPT and PT lies in the prompt initialization. We use Xavier uniform initialization to initialize two sets of prompt tokens: wavelet initial tokens $\mathbf{T}=\left\{\mathbf{T}_{k} \in \mathbb{R}^{w \times d} \mid k \in \mathbb{N}, 1 \leq k \leq l\right\}$ and prompt token
$\mathbf{P}=\left\{\mathbf{P}_{k} \in \mathbb{R}^{p \times d} \mid k \in \mathbb{N}, 1 \leq k \leq l\right\}$, where $w$ and $p$ denotes the preset number of Wavelet tokens and PT tokens, respectively. For the wavelet initial token, we use the efficient and straightforward wavelet Haar to perform the DWT transformation. Haar wavelets consist of the low-pass filter L, and the high-pass filter H, as follows:
\begin{equation}
	L=\frac{1}{\sqrt{2}}[1,1]^{T}, H=\frac{1}{\sqrt{2}}[1,-1]^{T}.
\end{equation}
We can obtain four sub-bands, which can be expressed as:
\begin{equation}
	T_{L L},\left\{T_{L H}, T_{H L}, T_{H H}\right\}=\operatorname{DWT}(T).
\end{equation}
The Haar wavelet transform generates four components: the low-frequency component (LL), as well as the high frequency in the vertical (LH), horizontal (HL), and diagonal (HH) directions. Each component has a size of $\frac{w}{2} \times \frac{d}{2}$, and then we reshape each component to a size of $\frac{w}{4} \times d$. Based on this operation, each token can correspond to a specific frequency component. Finally, we concatenate LL, LH, HL, and HH components to form the wavelet prompt $\mathbf{W}=\left\{\mathbf{W}_{k} \in \mathbb{R}^{w \times d} \mid k \in \mathbb{N}, 1 \leq k \leq l\right\}.$ The above process is illustrated in the lower part of Figure \ref{fig:pipeline}.

After obtaining the wavelet prompt, we concatenate it with the prompt token $P$ at each layer. Thus, the WPT process can be illustrated as follows:
\begin{equation}
	[Z_{i}, E_{i}] = \textcolor{blue}{{L}_{i}}([\textcolor{red}{{W}_{i}}, \textcolor{red}{{P}_{i}},E_{i-1}]),     \quad \text{for } i=1,2,...,l.
\end{equation}
Similar to PT-SSL-AASIST, the output of the transformer final layer $I = [Z_{l}, E_{l}]$ will be sent to the AASIST backend and trained using the WCE loss.

\section{All-Type ADD Benchmark}
In this section, we will present the benchmark experimental setup, including the four type ADD datasets used, the CMs employed, the training and testing protocols, and the detailed implementation of the entire experiment.

\subsection{Dataset}
To evaluate CM's ability to detect all types of deepfake audio, the selection of datasets is crucial. The principles for selection include being relatively clean and devoid of partially spoofed scenarios. Our aim is to thoroughly explore the capabilities of CMs in relatively clean environments, as removing other interferences such as noise is beneficial for studying cross-type ADD. Details of the dataset can be found in Table \ref{tab:dataset}.
\begin{table}[t]
	\centering
	\setlength{\tabcolsep}{2.5pt}
	\fontsize{8pt}{9pt}\selectfont
	
	\renewcommand{\arraystretch}{0.95}
	\begin{tabular}{c|c|ccc}
		\toprule
		Type &Source & Train & Dev & Eval \\
		\midrule
		Speech & 19LA & 25,380 & 24,844 & 71,237 \\
		Sound & Codecfake-A3 & 69,378 & 9,911 &19,823 \\
		Singing & CtrSVDD & 84,404	&43,625	&92,769 \\
		Music & FakeMusicCaps &20,861 &6,058	&6,122 \\
		\midrule
		All & Combined Sources & 199,023 & 84,438 & 189,951 \\
		\bottomrule
	\end{tabular}
	\caption{Statistics of all-type ADD benchmark in terms of training, development, and evaluation set.}
	\label{tab:dataset}
\end{table}

\textbf{Speech-19LA}. A widely used benchmark containing 12,456 real and 108,978 fake samples from 11 TTS and 8 VC systems (A01–A19), with A01–A06 for training and A07–A19 for evaluation, ensuring no overlap in spoofing methods between sets.

\textbf{Sound-Codecfake-A3}. We chose the Codecfake A3 subset for sound experiments. The real source domain is from the training subset of Audiocaps \cite{kim2019audiocaps}, and the fake sounds are generated using AudioGen based on the corresponding caption. This condition includes 49,274 real sounds and 49,838 fake sounds. We randomly divided all the sound data into training, validation, and evaluation sets in a ratio of 7:1:2.

\textbf{Singing voice-CtrSVDD}. SVDD \cite{zhang2024svdd} is the first singing voice detection challenge. From the SVDD challenge, built on Mandarin and Japanese singing datasets with 14 SVS/SVC systems. A01–A08 are used for training, A09–A14 for testing, following the original protocol.

\textbf{Music-FakeMusicCaps}. 
FakeMusicCaps \cite{comanducci2024fakemusiccaps} is a deepfake music detection dataset. The real source domain of FakeMusicCaps is the MusicCaps \cite{agostinelli2023musiclm} dataset, which consists of 5.5k 10-second music clips from AudioSet \cite{gemmeke2017audio}, each paired with an annotation by a professional musician. Built from MusicCaps as real source and SunoCaps-style captions to synthesize fake music using six methods (TTM01–TTM05 + unknown). Real clips are split 7:1:2, while fake music uses TTM01–TTM03 for training, TTM04 for validation, and TTM05 + unknown for testing.

\subsection{Baseline Countermeasure}

We establish five baseline models—Spec-Resnet, AASIST, and three SSL-enhanced variants: MERT-AASIST, WavLM-AASIST, and XLSR-AASIST—defined by their front-end and AASIST back-end combinations.

\textbf{Spec-Resnet} uses spectrograms with ResNet~\cite{he2016deep}, representing traditional feature-based methods. Though often outperformed by SSL features, its cross-type generalization merits further study.

\textbf{AASIST} is a strong deepfake detection model operating directly on waveforms. It extracts high-level features via sinc convolutions~\cite{ravanelli2018speaker} and residual blocks, followed by spectral-temporal attention for binary classification.

\textbf{SSL-AASIST} variants incorporate:
\begin{itemize}
	\item \textbf{MERT}, an SSL model designed for music understanding with strong performance in MIR tasks;
	\item \textbf{WavLM}, excelling in speech tasks and requiring further study in ADD scenarios;
	\item \textbf{XLSR}, widely recognized as the most effective SSL feature for cross-lingual and cross-type ADD tasks~\cite{phukan2024heterogeneity, pascu24_interspeech}.
\end{itemize}

We further explore four training paradigms for SSL-AASIST: 
\textbf{FR (frozen)} and \textbf{FT (fine-tuned)} differ by whether SSL parameters are updated during training; 
\textbf{PT} and \textbf{WPT}, our proposed methods, are described in Section~\ref{sec:cm}.

\subsection{Training and Evaluation Protocol}
To evaluate the all-type ADD capability of CM, we first conducted single-type training experiments, where the model was trained on one type of ADD dataset and tested on other types. In these experiments, the five CMs mentioned in the previous section were trained using a single type of training set and tested separately on each type of test set. For SSL-AASIST, different training paradigms can be employed, including FR, FT, PT, and WPT. To further address the all-type ADD task, we conducted all-type co-training experiments. Specifically, we trained the CM using all types of training set and tested it on each type of evaluation set.

\subsection{Implementation Details}

For the pre-processing of the ADD baseline models, all audio samples were first down-sampled to 16,000 Hz and trimmed or padded to 64600 samples (same as the original AASIST and SSL-AASIST). For the Spec-Resnet, the spectrogram was computed with the number of FFT points set to 512, the hop length set to 160, and the window length set to 512. The back-end Resnet used Resnet18 followed by a fully connected layer to down-sample to 2 dimensions. 
For the training paradigm, FT-SSL-AASIST adopted the training parameters from Tak et al. \cite{tak2022automatic}, with an initial learning rate of $10^{-6}$ and a batch size of 14. FR, PT, and WPT used a learning rate of $5 \times 10^{-4}$ and batch size 32. SSL features had shape (201, 1024) for 4s audio.
For single-type training, models trained for 50 epochs, halving the learning rate every 10 steps. Co-training ran for 20 epochs with LR halved every 4 steps.

\section{Experiments}

\subsection{Investigation for Single-Type Training}

\begin{table}[t]
	\centering
	\setlength{\tabcolsep}{2.5pt}
	\fontsize{8pt}{10pt}\selectfont
	\begin{tabular}{c|c|cccc|c}
		\toprule
		Train & Countermeasure & Speech & Sound & Singing & Music & AVG \\
		\midrule
		Speech & Spec-Resnet & 5.58 & 48.64 & 45.15 & 47.01 & 36.60 \\
		Speech & AASIST & 1.48 & 48.32 & 40.71 & 47.75 & 34.57 \\
		Speech & FR-MERT-AASIST & 4.80 & \textbf{47.60} & 44.51 & 48.89 & 36.45 \\
		Speech & FR-WavLM-AASIST & 2.49 & 47.96 & 38.67 & \textbf{42.75} & 32.97 \\
		Speech & FR-XLSR-AASIST & \textbf{1.28} & 49.51 & \textbf{29.72} & 49.82 & \textbf{32.58} \\
		\midrule
		Sound & Spec-Resnet & 49.67 & 8.87 & 47.77 & 44.22 & 37.63 \\
		Sound & AASIST & 37.39 & \textbf{0.43} & 42.56 & \textbf{10.44} & 22.71 \\
		Sound & FR-MERT-AASIST & 23.37 & 0.64 & 43.31 & 49.82 & 29.29 \\
		Sound & FR-WavLM-AASIST & 39.25 & 7.09 & 36.67 & 46.47 & 32.37 \\
		Sound & FR-XLSR-AASIST & \textbf{16.88} & 2.40 & \textbf{31.82} & 33.65 & \textbf{21.19} \\
		\midrule
		Singing & Spec-Resnet & 37.54 & 46.04 & 23.59 & \textbf{32.70} & 34.97 \\
		Singing & AASIST & 33.06 & 38.23 & 20.51 & 36.62 & 32.11 \\
		Singing & FR-MERT-AASIST & 43.86 & 42.88 & 29.95 & 44.24 & 40.23 \\
		Singing & FR-WavLM-AASIST & 16.19 & 41.80 & 18.74 & 39.18 & 28.98 \\
		Singing & FR-XLSR-AASIST & \textbf{12.89} & \textbf{34.41} & \textbf{9.45} & 35.87 & \textbf{23.16} \\
		\midrule
		Music & Spec-Resnet & 46.33 & 47.52 & 48.33 & 15.61 & 39.45 \\
		Music & AASIST & 31.81 & 47.26 & 44.12 & 8.36 & 32.89 \\
		Music & FR-MERT-AASIST & \textbf{27.88} & 44.45 & \textbf{34.56} & \textbf{7.62} & \textbf{28.63} \\
		Music & FR-WavLM-AASIST & 45.88 & 43.64 & 45.15 & 15.80 & 37.62 \\
		Music & FR-XLSR-AASIST & 48.89 & \textbf{40.54} & 43.41 & 9.67 & 35.63 \\
		\bottomrule
	\end{tabular}
	\caption{EER (\%) results of the countermeasures (frozen SSL) trained on single-type ADD training set.}
	\label{tab:singletypefrozen}
\end{table}

In this section, we trained deepfake countermeasures (CMs) using single-type datasets and evaluated Spec-Resnet, AASIST, and FR-SSL-AASIST, as shown in Table \ref{tab:singletypefrozen}. This setup allows us to assess the inherent detection capabilities of frozen SSL features.
For speech, XLSR-AASIST performed best, achieving the lowest in-domain (ID) EER (1.28\%) and average EER (32.58\%). It also showed reasonable transferability to singing (EER: 29.72\%)—much better than near-chance performance on sound and music—suggesting shared features between speech and singing.
For sound, AASIST achieved the best in-domain EER (0.4\%), likely due to overfitting on the single-method CodecFake-A3 data. Interestingly, this overfitting also benefitted music detection, hinting at similarities between non-speech audio types.
For singing, XLSR-AASIST again led with an ID EER of 9.45\% and average EER of 23.16\%, and even generalized well to speech (EER: 12.89\%), reinforcing the speech-singing similarity.
For music, MERT-AASIST achieved the best performance (ID: 7.62\%, Avg: 28.63\%), consistent with MERT’s strength in music representation.

\begin{table}[t]
	\centering
	\setlength{\tabcolsep}{1.5pt}
	\fontsize{8pt}{10pt}\selectfont
	\begin{tabular}{c|c|cccc|c}
		\toprule
		Train & Countermeasure & Speech & Sound & Singing & Music & AVG \\
		\midrule
		Speech & FT-Mert-AASIST & 6.99 & 48.37 & 48.86 & 44.43 & 37.16 \\
		Speech & FT-WavLM-AASIST & 1.50 & \textbf{44.62} & 35.77 & 42.19 & 31.02 \\
		Speech & FT-XLSR-AASIST & \textbf{0.38} & 49.57 & \textbf{29.76} & \textbf{31.01} & \textbf{27.68} \\
		\midrule
		Sound & FT-MERT-AASIST & 21.69 & \textbf{0.20} & 48.23 & 43.68 & 28.45 \\
		Sound & FT-WavLM-AASIST & 32.10 & \textbf{0.20} & 47.76 & \textbf{21.72} & 25.45 \\
		Sound & FT-XLSR-AASIST & \textbf{9.22} & 0.21 & \textbf{35.96} & 44.77 & \textbf{22.54} \\
		\midrule
		Singing & FT-MERT-AASIST & 43.51 & 41.92 & 30.58 & 41.81 & 39.46 \\
		Singing & FT-WavLM-AASIST & 13.07 & 36.68 & 8.00 & 40.32 & 24.52 \\
		Singing & FT-XLSR-AASIST & \textbf{7.56} & \textbf{31.08} & \textbf{5.60} & \textbf{37.36} & \textbf{20.40} \\
		\midrule
		Music & FT-MERT-AASIST & \textbf{24.03} & \textbf{46.50} & \textbf{44.79} & \textbf{15.53} & \textbf{32.21} \\
		Music & FT-WavLM-AASIST & 48.82 & 48.82 & 46.22 & 47.03 & 47.72 \\
		Music & FT-XLSR-AASIST & 39.03 & 47.99 & 47.93 & 48.70 & 45.91 \\
		\bottomrule
	\end{tabular}
	\caption{EER (\%) results of the countermeasures (finetuned SSL) trained on single-type ADD training set.}
	\label{tab:singletypefinetune}
\end{table}

Then, we investigate the SSL-AASIST through fine-tuning full SSL layer as shown in Table \ref{tab:singletypefinetune}. Overall, the final results were completely consistent with the frozen SSL models. For the speech-trained, sound-trained, and singing-trained CMs, the best performance was achieved by FT-XLSR-AASIST, with average EERs of 27.68\%, 22.54\%, and 20.40\%, respectively. For the music-trained CMs, the best performance was achieved by FT-MERT-AASIST, with an average EER of 32.21\%. It is also noteworthy that FT-SSL-AASIST consistently achieved lower EERs across various ID tasks compared to FR-SSL-AASIST. For instance, FT-XLSR-AASIST achieved EER of 0.38\% for speech, 0.21\% for sound, and 5.60\% for singing, representing reductions of 0.9\%, 2.19\%, and 3.85\% respectively compared to FR-SSL-AASIST. However, for music-trained SSL, all features exhibited a decline compared to FR, highlighting the challenges of fine-tuning. This indicates that fine-tuning, while requiring extensive parameter training, may also necessitate setting different hyper-parameters based on the type of data and SSL.

\begin{table}[t]
	\centering
	\setlength{\tabcolsep}{3pt}
	\begin{tabular}{c|c|cccc|c}
		\toprule
		Token & Parm & Speech & Sound & Singing & Music & AVG \\
		\midrule
		2  & 0.50M & 0.75  & 45.29 & 35.00 & 42.71 & 30.94 \\
		10 & 0.69M & \bf 0.22  & 47.26 & \bf 33.84 & 41.85 &\bf  30.79 \\
		20 & 0.94M & 0.58  & 44.11 & 43.35 & 41.64 & 32.42 \\
		100 & 2.90M & 3.01  & \bf 37.05 & 49.41 & \bf 35.66 & 31.28 \\
		200 & 5.36M & 4.99  & 44.45 & 47.61 & 36.37 & 33.36 \\
		\bottomrule
	\end{tabular}
	\caption{EER (\%) comparison with different number of token.}
	\label{tab:token}
\end{table}
\begin{table}[t]
	\centering

	\setlength{\tabcolsep}{2pt}
	\begin{tabular}{c|cccc|c}
		\toprule
		Paradigm &Speech & Sound & Singing & Music &AVG \\
		\midrule
		Shallow-PT & 0.75	&\bf 45.29	&39.87	&44.24	&32.54 \\
		After-PT   & 0.53	&46.88	&41.55	&44.05	&33.25 \\
		Del-PT     & 0.72	&47.23	&41.45	&42.87	&33.07 \\
		PT         & \bf 0.22  & 47.26 & \bf 33.84 & \bf 41.85  &\bf 30.79 \\
		\bottomrule
	\end{tabular}
	\caption{EER (\%) comparison with different paradigms.}
	\label{tab:connect}
\end{table}
\begin{figure}[!t]
	\centering
	\subfloat{\includegraphics[width=3in]{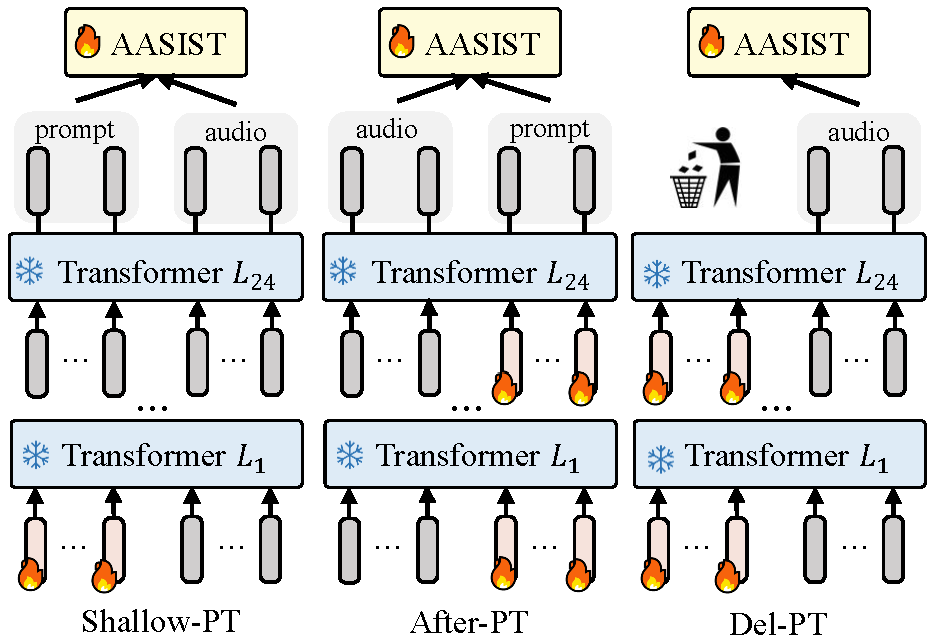}}
	\hfil
	\caption{Different paradigms of PT-SSL-AASIST.}
	\label{fig:paradigms} 
\end{figure}

\subsection{Prompt Tuning Countermeasures} 
\textbf{PT-SSL-AASIST.} To evaluate the effectiveness of PT and determine their optimal parameters, we integrated PT into the XLSR-AASIST, which performed best in the previous section, trained on the speech dataset. There are two aspects worth investigating for PT: the preset number of tokens for PT and the paradigm for PT (connection method, prompt position, etc.).
We conducted ablation experiments on the number of tokens and the paradigm, as shown in the Table \ref{tab:token} and Table \ref{tab:connect}, respectively. 

For the number of tokens in PT, we experimented with 2, 10, 20, 100, and 200. The results showed that when the token number was set to 10, the best speech test set EER of 0.22\% and the lowest average EER of 30.79\% were achieved. As the number of tokens increased, the parameter count for PT training also increased, but the effectiveness decreased. This is may due to the fact that the number of audio tokens is 201, and an excessive number of prompt tokens can cause the audio tokens to become sparse, hindering the learning of the audio's inherent information.

After determining the number of tokens, we investigated three paradigms for PT-SSL-AASIST, including Shallow-PT, After-PT, and Del-PT, as shown in Figure \ref{fig:paradigms}. Shallow-PT refers to inserting learnable prompts only in the first transformer encoder layer, which can demonstrate the importance of the deep paradigm where prompts are inserted in each layer. After-PT places the prompt position after the audio token, which might be effective due to the artifact information located in the silent region at the beginning of the audio \cite{zhang2023impact}. Del-PT involves deleting the prompt token in the last layer, using only the audio tokens for classification, a method considered effective in some vision tasks \cite{darcet2024vision}. Experimental results indicate that our proposed PT-SSL-AASIST paradigm is optimal, where prompts are inserted in each layer and the final layer combines the prompt and audio tokens for input into AASIST.
\begin{table}[t]
	\centering

	\setlength{\tabcolsep}{3pt}
	\renewcommand{\arraystretch}{0.95}
	\fontsize{8pt}{10pt}\selectfont
	\begin{tabular}{c|c|cccc|c}
		\toprule
		WPTs& PTs& Speech & Sound & Singing & Music & AVG \\
		\midrule
		0  & 10 & 0.22 & 47.26 & 33.84 & 41.85 & 30.79 \\
		2  & 8  & 0.16 & 49.18 & 38.22 & 34.84 & 30.60 \\
		4  & 6  & 0.15 & \textbf{45.36} & \textbf{33.32} & \textbf{28.61} & \textbf{26.86} \\
		6  & 4  & 0.16	&47.86	&36.21	&31.52	&28.94
		\\
		8  & 2  & 0.18 & 47.40 &40.68	&32.25	&30.13
		\\
		10 & 0  & \textbf{0.11} & 49.40 & 36.97 & 43.68 & 32.54 \\
		\bottomrule
	\end{tabular}
	\caption{EER (\%) comparison with different number of wavelet prompt tokens (WPTs) and prompt tokens (PTs).}
	\label{tab:wtoken}
	
\end{table}
\begin{table}[t]
	\centering
	\setlength{\tabcolsep}{2pt}
	\fontsize{8pt}{10pt}\selectfont
	\begin{tabular}{c|c|cccc|c}
		\toprule
		Countermeasure & Parm & Speech & Sound & Singing & Music & AVG \\
		\midrule
		FR-XLSR-AASIST & 0.45M & 1.28 & 49.51 & \textbf{29.72} & 49.82 & 32.58 \\
		FT-XLSR-AASIST & 315.89M& 0.38 & 49.57 & 29.76 & 31.01 & 27.68 \\
		PT-XLSR-AASIST & 0.69M & 0.22 & 47.26 & 33.84 & 41.85 & 30.79 \\
		WPT-XLSR-AASIST & 0.69M & \textbf{0.15} & \textbf{45.36} & 33.32 & \textbf{28.61} & \textbf{26.86} \\
		\bottomrule
	\end{tabular}
	\caption{EER (\%) and training parameters comparison with different paradigms of speech-trained XLSR-AASIST.}
	\label{tab:comparessl}
\end{table}
\begin{figure}[!t]
	\centering
	\subfloat{\includegraphics[width=3.3in, height=1.6in]{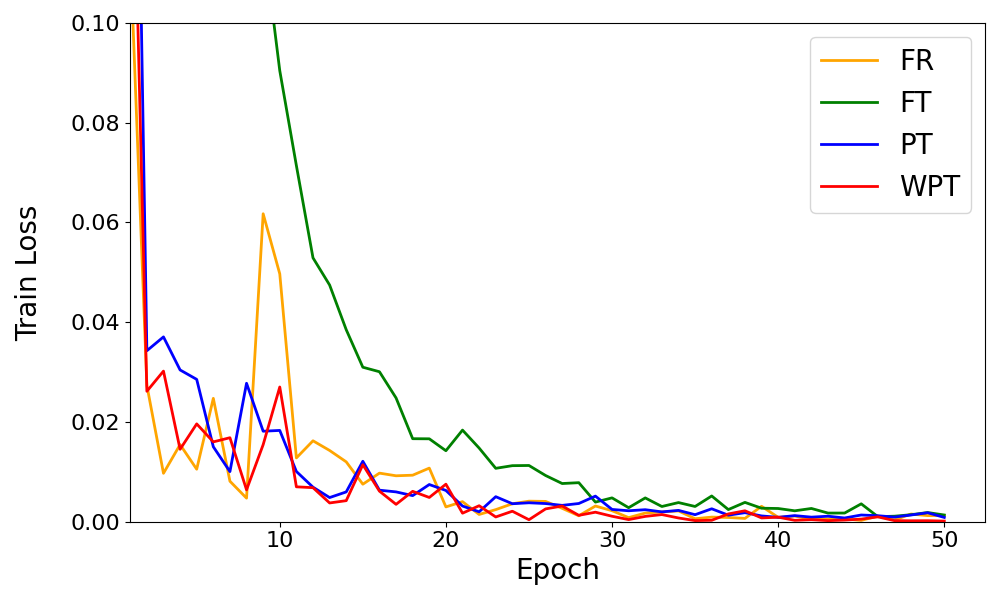}}
	\hfil
	\caption{Convergence speed of different paradigms.}
	\label{fig:trainloss} 
\end{figure}
\begin{table}[t]
	\centering
	\setlength{\tabcolsep}{2pt}
	\renewcommand{\arraystretch}{0.85}
	\fontsize{8pt}{10pt}\selectfont
	\begin{tabular}{c|cccc|c}
		\toprule
		Countermeasure & Speech & Sound & Singing & Music & AVG \\
		\midrule
		Spec-Resnet & 29.37 & 23.37 & 37.17 & 42.75 & 33.17 \\
		AASIST & 3.78 & 0.86 & 20.01 & 11.70 & 9.09 \\
		\midrule
		FR-WavLM-AASIST & 3.44 & 10.21 & 17.83 & 26.02 & 14.38 \\
		FT-WavLM-AASIST & \textbf{1.31} & 2.53 & 16.48 & 22.90 & 10.81 \\
		PT-WavLM-AASIST & 3.09 & 8.81 & 15.84 & \textbf{16.73} & 11.12 \\
		WPT-WavLM-AASIST & 2.04 & \textbf{1.10} & \textbf{9.28} & 18.21 & \textbf{7.66} \\
		\midrule
		FR-MERT-AASIST & \textbf{2.90} & 4.60 & \textbf{12.14} & 24.91 & 11.14 \\
		FT-MERT-AASIST & 6.24 & 1.17 & 31.67 & 13.77 & 13.21 \\
		PT-MERT-AASIST & 6.06 & 1.28 & 32.59 & 9.29 & 12.31 \\
		WPT-MERT-AASIST & 6.59 & \textbf{1.01}& 22.68 & \textbf{8.53} & \textbf{9.70} \\
		\midrule
		FR-XLSR-AASIST & 3.02 & 5.45 & 10.86 & 22.67 & 10.50 \\
		FT-XLSR-AASIST & 1.77 & \textbf{0.49} & 8.93 & 8.71 & 4.98 \\
		PT-XLSR-AASIST & 2.00 & 1.11 & 14.54 & 9.29 & 6.74 \\
		WPT-XLSR-AASIST & \textbf{0.72} & 1.29 & \textbf{7.47} & \textbf{4.83} & \textbf{3.58} \\
		\bottomrule
	\end{tabular}
	\caption{EER (\%) results for the countermeasures co-trained on the complete ADD training set.}
	\label{tab:cotrained}
\end{table}

\textbf{WPT-SSL-AASIST.} After deciding the PT architecture, we introduced WPT, applying DWT to a part of the ten prompt tokens to better capture the frequency information of the audio. We first investigate the optimal number of tokens by comparing different settings of wavelet prompt tokens (WPTs) and standard prompt tokens (PTs). As shown in Table \ref{tab:wtoken}, the best overall performance is achieved when WPTs = 4, yielding an average EER of 26.86\%. Interestingly, this configuration also naturally aligns each of the four frequency bands (LL, LH, HL, HH) with a single token. Although the configuration with WPTs = 10 achieves the lowest EER of 0.11\% on the speech type, its performance degrades significantly on other types, making it less favorable overall. Based on the optimal performance observed, we adopt 4 WPTs for the remainder of our experiments.
\begin{figure}[!t]
	\centering
	\subfloat{\includegraphics[width=3.3in]{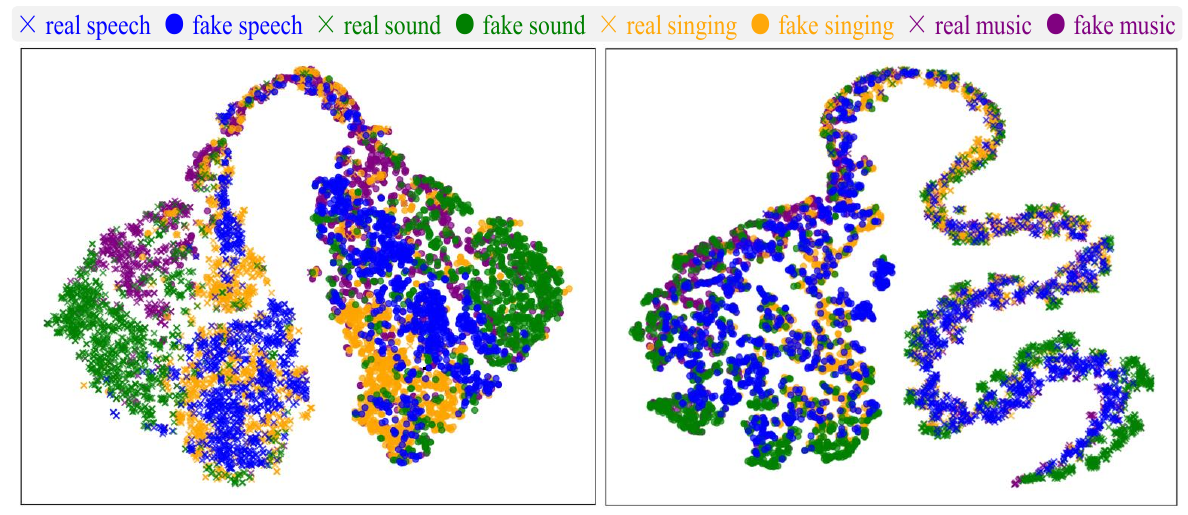}}
	\hfil
	\caption{T-SNE visualization for FT-XLSR-AASIST (left) and WPT-XLSR-AASIST (right). Different colors indicate features from different types: blue=speech, green=sound, orange=singing, purple=music. Different shapes represent different categories: cross=real, point=fake.}
	\label{fig:tsne} 
\end{figure}

\begin{figure*}[!t]
	\centering
	\subfloat{\includegraphics[width=6.4in]{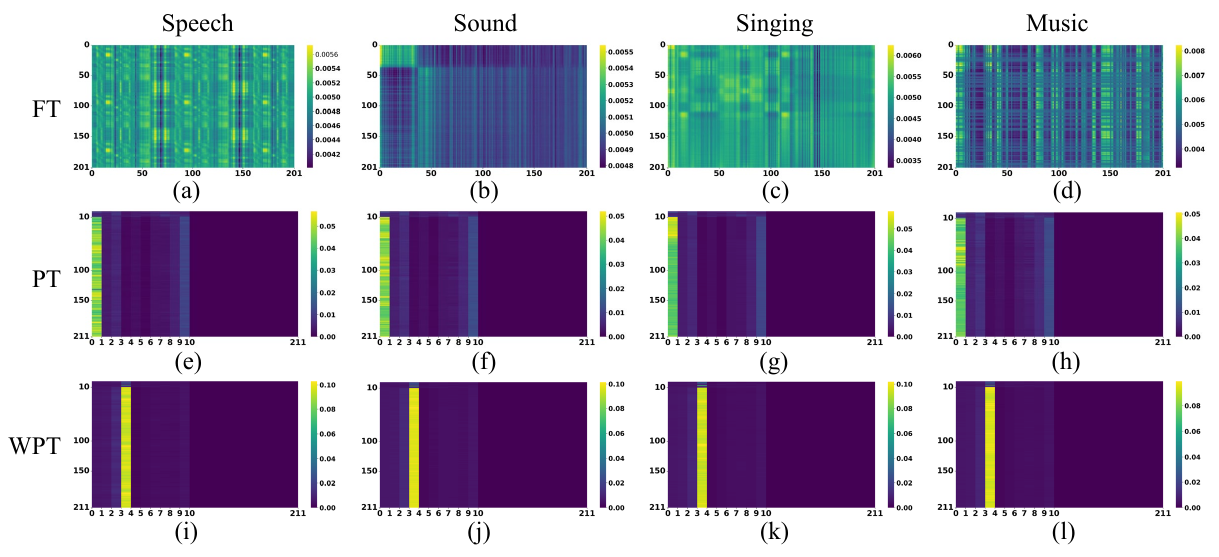}}
	\hfil
	\caption{Attention map of the final transformer in co-trained XLSR-AASIST. Each column corresponds to the same deepfake audio sample. For both PT and WPT, we magnified the position of prompt tokens (1-10).}
	\label{fig:attention} 
\end{figure*}

Then, we compared the performance of FR, FT, PT, and WPT using the speech-trained XLSR-AASIST, as shown in Table \ref{tab:comparessl}. It can be observed that WPT achieved the best results compared to PT, obtaining a 0.15\% EER on the ID speech evaluation set, with an average EER of 27.55\%. Moreover, WPT does not increase the number of training parameters compared to PT, and compared to FT, the training parameters are reduced by 458 times. Overall, WPT outperformed FT, which in turn outperformed PT, and PT outperformed FR. We also recorded the training convergence speeds, as shown in Figure \ref{fig:trainloss}. It can be observed that FR, PT, and WPT converged significantly faster than FT, and both FT and PT showed less fluctuation compared to FR during convergence.

\subsection{Co-trained Countermeasures}

Although the speech-trained WPT-XLSR-AASIST achieved extremely low EER on speech deepfake test set, it still exhibited significant performance degradation on detecting deepfake sound, singing, and music. Therefore, we began to investigate co-trained CM, combining the training sets of the four types to achieve all-type audio deepfake CM.

The results of the co-training experiment are shown in Table \ref{tab:cotrained}. Firstly, the effectiveness of the data-driven approach can be observed, with a significant reduction in average EER compared to single-type trained CMs. The best performing SSL in the co-training experiment is the XLSR-AASIST. For the XLSR-AASIST training paradigm, WPT outperformed FT, PT, and FR, achieving EERs of 3.58\%, 4.98\%, 6.74\%, and 10.50\%, respectively. This training paradigm's performance aligns with that of the speech-trained XLSR-AASIST shown in Table \ref{tab:comparessl}. Notably, WPT consistently achieves the best performance across different SSL features. For instance, WPT-WavLM-AASIST and WPT-MERT-AASIST achieve EER of 7.66\% and 9.70\%, respectively.

\subsection{Interpretability}
\textbf{Type Invariance in T-SNE Visualization.} To further understand the interpretability of the WPT training paradigm, we first performed T-SNE visualization on the embeddings before the final fully connected layer of AASIST. Specifically, we applied T-SNE visualization to the embeddings from the co-trained FT-XLSR-AASIST and WPT-XLSR-AASIST on evaluation sets of four audio types. For each type, we selected 2,000 samples randomly, comprising 1,000 genuine samples and 1,000 fake samples. The results are presented in Figure \ref{fig:tsne}. Firstly, it can be observed that both FT and WPT are capable of separating the test real and fake samples. However, there is a notable difference. FT demonstrates distinct clustering within both the genuine and fake regions, where speech, sound, singing, and music samples form separate clusters. In contrast, WPT does not exhibit such separation within either the genuine or fake regions, resulting in overlap among the four types. This indicates that WPT maintains type invariance when performing the all-type ADD task.

\textbf{Type Invariance in Attention Distribution.} To further investigate the intrinsic differences in training paradigms for detecting deepfakes, we plotted the attention maps of the final transformer encoder layer, as shown in Figure \ref{fig:attention}. It is evident that FT exhibits different attention distributions when processing different types of audio. Interestingly, the attention patterns for speech and singing are similar, exhibiting overall high values with some regions of exceptionally high intensity. The attention patterns for sound and music are also similar, displaying a mix of high and low values in all region. This observation is consistent with the experimental results from single-type training. For the PT and WPT paradigms, we can observe consistency in their detecting of different types. The PT paradigm focuses on the first prompt token, but the values are not high, and there are noticeable value changes when dealing with different types, with some attention also present on the 10th prompt token. In contrast, WPT paradigm demonstrates significant invariance in detecting diverse audio types, with a focus on the 4th token corresponding to the wavelet HH token, which determines high-frequency details through diagonal orientation analysis.

\subsection{Conclusion}
In this paper, we are dedicated to studying the all-type ADD task. We are the first to establish a comprehensive benchmark for evaluating the performance of current CMs on the all-type ADD task. To achieve all-type CM, we propose the WPT-SSL training paradigm, which leverages wavelet prompts to capture the type-invariant auditory deepfake information of SSL features. Our proposed co-trained WPT-XLSR-AASIST achieves an average EER of 3.58\% across all-type ADD evaluation set. 

\section{Acknowledgements}
This work is supported by the Beijing Natural Science Foundation (No. L252143), and the National Natural Science Foundation of China (NSFC) (No.62201571, No.62101553, No.U21B20210).
\bibliography{myrefs}

\end{document}